\documentclass[prd,a4paper,twocolumn,preprintnumbers,nofootinbib]{revtex4}
\usepackage[a4paper, hdivide={1.91cm,,1.165cm}, vdivide={1.83cm,,3.8cm}]{geometry}

\usepackage{color}
\usepackage[hyperfootnotes=false]{hyperref}
\usepackage{youngtab}
\usepackage[english]{babel} 
\usepackage{amsmath}
\usepackage{float}
\usepackage{amssymb}
\usepackage{braket}
\usepackage{amsfonts}
\usepackage{mathrsfs}
\usepackage{enumerate}
\usepackage{graphicx}
\usepackage{array}
\usepackage{booktabs}
\usepackage{units}
\usepackage{textcomp}
\usepackage{mathtools}
\usepackage{cancel}
\usepackage{slashed}
\usepackage{multirow}
\usepackage{wasysym}
\usepackage{slashed}
\usepackage{comment}
\usepackage{relsize}
\usepackage{bm}

\usepackage{epstopdf}
\usepackage{caption}

\newcommand{\dd}{\text{d}}

\newcommand{\pts}[1]{\phantom{.}\hfill(\textit{#1~point}\ifthenelse{\equal{#1}{1}}{}{\textit{s}})}

\newcommand{\Msun}{{\ifmmode{{\rm{M_{\odot}}}}\else{${\rm{M_{\odot}}}$}\fi}}

\newcommand{\beq}{\begin{equation}}
\newcommand{\eeq}{\end{equation}}
\newcommand{\bea}{\begin{eqnarray}}
\newcommand{\ena}{\end{eqnarray}}

\newcommand{\lsim}{\mathrel{\mathop{\kern 0pt \rlap
{\raise.2ex\hbox{$<$}}}
\lower.9ex\hbox{\kern-.190em $\sim$}}}
\newcommand{\gsim}{\mathrel{\mathop{\kern 0pt \rlap
{\raise.2ex\hbox{$>$}}}
\lower.9ex\hbox{\kern-.190em $\sim$}}}
\begin{document}

\preprint{ULB-TH/19-10}

\title{Constraints on Dark Matter from the Moon}

\author{Raghuveer Garani}
\email{rgaranir@ulb.ac.be}
\affiliation{Service de Physique Th\'eorique, Universit\'e Libre de Bruxelles, Boulevard du Triomphe, CP225, 1050 Brussels, Belgium}

\author{Peter Tinyakov}
\email{ptinyakov@ulb.ac.be}
\affiliation{Service de Physique Th\'eorique, Universit\'e Libre de Bruxelles, Boulevard du Triomphe, CP225, 1050 Brussels, Belgium}


\begin{abstract}

New and complimentary constraints are placed on the spin-independent interactions of dark matter with baryonic matter. Similar to the Earth and other planets, the Moon does not have any major internal heat source.  We derive constraints by comparing the rate of energy deposit by dark matter annihilations in the Moon to 12 mW/m$^2$ as measured by the Apollo mission. 
For light dark matter of mass $\mathcal{O}(10)$ GeV, we also examine the possibility of dark matter annihilations in the Moon limb. In this case, we place constraints by comparing the photon flux from such annihilations to that of the Fermi-LAT measurement of $10^{-4}$ MeV/cm$^2$s. This analysis excludes spin independent cross section $\gtrsim 10^{-37}$ $\rm{cm}^2$ for dark matter mass between 30 and 50 GeV.  

\end{abstract}
\maketitle

\section{Introduction}
\label{sec:introduction}

A popular assumption about the dark matter (DM) is that it is composed of new
particles. While these particles have not been detected directly, there is an
ongoing campaign to constrain their properties from various arguments. From
cosmological observations DM particles must be sufficiently ``cold'' to
cluster at galactic and larger scales; their interaction with baryons and
electrons are constrained by the direct searches; their lifetime and
annihilation cross section must be compatible with the measured DM abundance
and must satisfy the constraints from the indirect detection
experiments~\cite{Tanabashi:2018oca}. While many regions of the parameter
space are now excluded, large areas still remain open, and new ways to
constrain them are therefore worth exploring.

If the DM particles can annihilate into the Standard Model (SM) particles, they
are typically expected to also scatter on ordinary matter. They may in this case accumulate in
astrophysical objects such as stars and planets, where their annihilation may
produce detectable effects thus allowing for the constraints on the DM
parameters to be placed. For instance, the DM accumulation in the Sun can lead
to a detectable neutrino
flux~\cite{Silk:1985ax,Srednicki:1986vj,Gould:1987ju,Gould:1989tu}.
If astrophysical object has no internal heat sources, the heat from the DM
annihilations can be detectable in some cases. The DM annihilation in neutron
stars~\cite{Kouvaris:2007ay,Bertone:2007ae,Kouvaris:2010vv,Baryakhtar:2017dbj,Chen:2018ohx,Garani:2019fpa,Acevedo:2019agu}, in white dwarfs~\cite{McCullough:2010ai,Bertone:2007ae}, 
in the Earth ~\cite{Mack:2007xj,Mack:2012ju,Kavanagh:2017cru}, and in Mars~\cite{Bramante:2019fhi} has been previously considered in this context.

In this paper we examine the observational consequences of DM accumulation and
annihilation in the Moon. One of such consequences is Moon heating. Like in Earth and other planets, internal heat sources of the Moon are several orders of magnitude less efficient than typical fusion reactions in stars. The source of internal heat in planetary bodies is attributed mainly to residual heat from gravitational energy left over from planetary accretion and core formation, and  the decay of radiogenic isotopes~\cite{10.3389/feart.2016.00046}. However, since the Moon is geologically inactive today, it is suggested that much of the Moon's total internal heat comes from continuous radiogenic decay~\cite{1976LPSC....7.3461T} (see Sect.~\ref{sec:moon-heat-flow} for details about the Moon composition and internal heat). This makes even a
small heat produced by DM annihilations potentially detectable. Although Moon
models are currently less advanced than the Earth ones, even the present-day
Moon heat flow measurements translate into constraints that are competitive
with those derived from the Earth, while its proximity and potential
accessibility to direct measurements promise significant improvement in the
future. In addition, Moon is believed to have lighter composition than Earth,
so the DM capture rate (and, consequently, the constraints) which depends on
the chemical composition in a resonant way is maximized at smaller DM masses,
thus some complementarity with the Earth constraints is expected.

Another potentially observable effect that we consider is the $\gamma$-ray
production. In some limited range of DM masses around few tens of GeV, a
sizable fraction of DM annihilations happens outside of the Moon. Such
annihilations are directly observable by the $\gamma$-ray telescopes. Due to
observational conditions, this effect is more important for the Moon than the
Earth. 

The constrains that we obtain are competitive with those previously derived
for the Earth, but are inferior to the direct detection constraints in the
same range of masses. The latter, however, do not cover the region of large
cross sections where the DM particles do not reach the underground detectors.
Like the Earth constraints, the Moon constraints are free from this problem.

Before going into a detailed calculation it is instructive to make
rough estimates. The rate at which the DM of mass $m$ crosses the Moon surface (the
geometric capture rate) is given by $\pi R^2_{\leftmoon} \rho_{\rm DM} v/m$, where
$R_{\leftmoon}=1737$~km is the Moon radius, while $\rho_{\rm DM}=0.3\,{\rm GeV/cm}^3$ and $v=270$~km/s are the DM density and velocity, assuming the standard DM halo
parameters at the Earth location in the Galaxy. If all this DM gets captured
and annihilates in the Moon producing heat, the total released power would be
$\sim 100$~TW.  This exceeds by a factor $\sim 200$ the measured value
$0.45$~TW (see Sect.~\ref{sec:moon-heat-flow} for details about the Moon). Requiring that the actual capture rate is smaller by the factor $1/200$ than this geometric
maximum gives constraints on the DM parameters. Likewise, the $\gamma$-ray
flux from the Moon is measured by the Fermi-LAT satellite to be $\sim
10^{-4}$MeV/cm$^2$s~\cite{Cerutti:2016gts}, which corresponds to
the total power of $\sim 3\times 10^{-7}$~TW. Again, requiring that the limb
of DM annihilations around Moon does not over-shine this measured value gives
additional constraints on DM parameters.

The rest of this paper is organized as follows. In
Sect.~\ref{sec:moon-heat-flow} we overview the Moon heat flow measurements and
the elemental composition of the moon.  In
Sect.~\ref{sec:capture-annihilation} we calculate the key quantities: the DM
capture rate, the evaporation rate, and the annihilation rate. In Sect.~\ref{sec:results} we present the exclusion curves for spin-independent interactions of dark matter with protons. Finally, in Sect.~\ref{sec:concl} we present concluding remarks.

\section{Moon composition and heat flow measurements}
\label{sec:moon-heat-flow}
Most of our knowledge about the internal structure of the Moon and the
physical properties of its different layers comes from seismological
data~\cite{1982LPSC...13..117N}. Similar to studies of the Earths' core, the
determination of the density distribution of the Moon from bulk sound velocity
of seismic waves in combination with normal modes is a well-established method
with statistical uncertainties in the mantle at the several percent level, and
larger errors for core densities. For the purpose of this study we
consider two benchmark models that correspond to maximal (MAX) and minimal (MIN) core density. We present results in Sect.~\ref{sec:results} for both the above models. We consider the model MIN to be more conservative since it corresponds to Very Preliminary Reference Moon Model (VPRMM)~\cite{GARCIA201196}. Below we summarize the current understanding of the composition of the moon and its geological layers, and more importantly the internal heat flux of the moon. Finally, we list all relevant information in table ~\ref{tab:abundance-max} (\ref{tab:abundance-min}) for model MAX (MIN).

{\it Lunar Core}: Information about the size of the {\it lunar core} is
derived from reprocessing of Apollo-era seismic
analyses~\cite{Weber309}. According to the review Ref.~\cite{Wieczorek2006},
existence of a lunar core of a radius $250-450$~km is suggested. However,
there is a debate on the composition of the core. Earlier works have shown
that the core could be composed of Fe, FeS, FeS-Fe or Fe-FeS-C alloy, or dense
Ti-rich silicate of density $\sim 8\,{\rm g/cm^3}$, but more data is required
to discriminate between them. Note that inferences about core size is highly
dependent on the modeled core composition~\cite{Wieczorek2006}. Also note
that there is a ``trade-off'' between the sulfur content of the lunar core and
its temperature. For instance, a 10\% increase of sulfur content decreases the
core temperature by 600~K~\cite{Weber309}. For the benchmark model MAX we adopt the value of core radius to be 450~km with core density $9\,{\rm g/cm^3}$. For the benchmark model MIN we consider the value of core radius to be 380~km and core density $5\,{\rm g/cm^3}$~\cite{GARCIA201196}. For both the cases above we assume a sulfur composition of 5\% by weight and core temperature 1700~K. Other parameters are adjusted to reproduce the correct mass and radius of the Moon.

{\it Lunar Mantle}: As samples of the lunar mantle have not been identified
within the lunar sample collection, information concerning its composition and
structure can only come from indirect sources, such as the analyses of the
mare basalts and volcanic glasses, and the Apollo seismic data.
Investigations of thermodynamically stable
mineral phases that are consistent with the Apollo seismic data indicate that
the upper ~500 km of the mantle is predominantly composed of orthopyroxene,
with smaller abundances of olivine, clinopyroxene, plagioclase and garnet. In
contrast, the lower mantle is predominately composed of olivine, with lesser
quantities of garnet and possibly
clinopyroxene~\cite{Wieczorek2006,1991lsug.book.....H}. In this work we assume
the lunar mantle extends up to 1180 km (1317 km) from the core, and has an average density of $\sim 3.2 \,{\rm g/cm^3}$ ($\sim 3.5 \,{\rm g/cm^3}$) for the benchmark model MAX (MIN), respectively.

{\it Lunar Crust}: Reanalyses of the Apollo seismic data by
Refs.~\cite{2002LPI....33.1548K,Lognonne2003} imply that the crust beneath the
Apollo 12 and 14 sites is between 27 and 50 km thick. We assume that the crust is 57 km (40 km) thick with an average density $\sim 2.9 \,{\rm g/cm^3}$ ($\sim 2.6 \,{\rm g/cm^3}$) for the benchmark model MAX (MIN), respectively.

{\it Elemental Composition}: All the chemical elements that make up the Earth
are also found on the Moon. On scales both large and small, however, the
abundances and distributions of the elements differ greatly between the two
bodies.  The Moon lacks a large iron core and is geologically inactive
today. It has no appreciable atmosphere and therefore does not undergo
chemical weathering of the type found on planets with
atmospheres~\cite{1991lsug.book.....H}. In Tabs.~\ref{tab:abundance-max} and  ~\ref{tab:abundance-min}, we merely list the elements and their concentrations adopted in the present study.
We have assumed that the elemental composition is the same for models MAX and MIN. We also assume that the mantle composition is the same as that of the crust. Uncertainties up to factor 2 are expected in the mass fraction quoted in tables. Further discussion is beyond the scope of this
paper.

\begin{table}[ht]
\centering
\begin{tabular}{|c| c| c|c|}
\hline
Radius  & Density  & Mass faction & number of nuclei \\
  {[km]} & [${\rm g/cm^3}$] & [\%] & [$\times 10^{46}$]\\ \hline
  0-450  &  9 & Fe (95)& Fe (3.52) \\
    &  &    S (5) & S (0.32) \\ \hline 
  450-1680 &  3.2 &  Fe O (18)  & O (88.0)\\
  & & Si$_2$ O$_3$ (45) & Mg (8.3) \\ 
  & & Al$_2$O$_3$ (14) & Al (10.3) \\ 
  & &  MgO (9)& Si (32.5) \\ 
  & &  CaO (12) & Ca (6.02) \\ 
   & & & Fe (9.4) \\ \hline
  1680-1737 &  2.9  & same & scaled by 0.097 \\\hline
\end{tabular}
  \caption{Moon layers and elemental abundances adopted in this work for the benchmark model MAX.}
\label{tab:abundance-max}
\end{table}

\begin{table}[ht]
\centering
\begin{tabular}{|c| c| c|c|}
\hline
Radius  & Density  & Mass faction & number of nuclei \\
  {[km]} & [${\rm g/cm^3}$] & [\%] & [$\times 10^{46}$]\\ \hline
  0-380  &  5 & Fe (95)& Fe (1.1) \\
    &  &    S (5) & S (0.10) \\ \hline 
  380-1697 &  3.4 &  Fe O (18)  & O (97.1)\\
  & & Si$_2$ O$_3$ (45) & Mg (9.2) \\ 
  & & Al$_2$O$_3$ (14) & Al (11.3) \\ 
  & &  MgO (9)& Si (35.8) \\ 
  & &  CaO (12) & Ca (6.6) \\ 
   & & & Fe (10.3) \\ \hline
  1697-1737 &  2.4  & same & scaled by 0.051 \\\hline
\end{tabular}
  \caption{Moon layers and elemental abundances adopted in this work for the benchmark model MIN.}
\label{tab:abundance-min}
\end{table}

{\it Moon Internal Heat Flow}: Direct measurements of surface heat flux of
the moon were carried out at Apollo sites 15 and
17~\cite{Grott2017InSH,doi:10.1002/2013JE004453,doi:10.1029/2018JE005579,doi:10.1002/jgre.20103,6612cdc0432a11df937d000ea68e967b}. According
to Ref.~\cite{6612cdc0432a11df937d000ea68e967b,doi:10.1029/JB092iB05p03453}
the two lunar heat flow determinations were made along Lunar highlands/mare
boundaries. Since the upper 2-3 km of the crust (megaregolith) is thin in mare
regions, heat passes easier through them than through the highland regions
(and even flows laterally from the highland toward the mare). Hence, the heat
flow is expected to be high along the boundary between highland and mare
regions. Upon adjusting the Apollo 17 heat flow for the boundary effect and assuming that the mean
megaregolith thickness is 2 km globally, a heat flow of $12\,{\rm mW/m^2}$
(0.45 TW total) is reported~\cite{doi:10.1029/JB092iB05p03453}. 
The heat flow measured at the Apollo sites is consistent, for a steady state model, with bulk uranium content of $\sim 40\times 10^{-9}$ parts per billion. A similar estimate is obtained from geochemical balance considerations, from correlations among refractory elements ratios~\cite{1976LPSC....7.3461T}.
 
{\it Future prospects}: There are several planned missions to the moon in the
coming decade. Some of these missions are particularly focused on landing a
rover on the moon for further geological exploration and mineralogy. It is
expected that such exploration would significantly reduce the uncertainties
in heat flow measurements and surface composition of the moon.

\section{Capture, Annihilation and Evaporation}
\label{sec:capture-annihilation}

The captured DM particles thermalize through collisions with the Moon nuclei
and form a cloud in the gravitational potential of the Moon. The size of this
cloud is determined by the average Moon temperature and the DM mass, and is
smaller for larger masses. When it is much smaller than the size of the Moon
(at large DM masses exceeding few tens of GeV), the evaporation can be
neglected and the annihilation rate equals the capture rate. When the DM cloud
is much larger than the size of the Moon (small DM masses) the evaporation
becomes dominant over annihilation, and the latter is only a small fraction of
the capture rate. In the intermediate case when the size of the DM cloud is
comparable to the size of the Moon, the annihilation is still not suppressed,
but a sizable fraction of particles annihilate {\em outside} the Moon, giving
rise to the observable $\gamma$-ray flux.

The differential equation governing the time evolution of the number of DM
particles gravitationally bound to the Moon $N$ is
\begin{equation}
  \frac{\dd N}{\dd t} = C - A N^2 - F N,
 \label{eq:equilibrium}
\end{equation}
where $C$ is the DM capture rate by the Moon, while second and third terms on
the right hand side represent the annihilation and evaporation rates. When
equilibrium is reached $dN/dt=0$, and the capture rate equals the sum of
annihilation and evaporation. The relative strength of these rates is
determined by the single combination of parameters, as in the equilibrium
one has
\begin{eqnarray}
\nonumber
FN  &=& {2C\over x} \left( \sqrt{x+1} -1 \right),\\
\nonumber
AN^2 &=& C - FN, 
\end{eqnarray}
where
\[
x= 4CA/ F^2.
\]
Evaporation is important at $x\ll 1$ and negligible in the opposite limit, in
which case the annihilation rate becomes equal to the capture rate.

We now calculate the three terms in Eq.~(\ref{eq:equilibrium}). Consider the
capture rate $C$ first. It depends on the Moon composition which we assume to
consist of several species with number densities $n_i(r)$ different in
different layers as given in Tabs.~\ref{tab:abundance-max} and~\ref{tab:abundance-min}, and masses
$m_i$. We take into account the motion of the Moon ($u=220$~km/s) with respect to the DM distribution, and assume the latter to be Maxwellian
  with the velocity dispersion $v_d=270$~km/s in the galactic rest frame. The capture rate is~\cite{Gould:1987ir,Gould:1999je}
\begin{eqnarray}
  C &=& \sum_i \sqrt{{6\over \pi}} {\rho_{\rm DM} \over m\, v_d}
  \int dV v_e^2 \sigma_i n_i  \times  \nonumber \\
  & & \frac{1}{2 \eta A^2_i} \left[ \left(A_{i,+ } A_{i,- } - \frac{1}{2}\right)\left(\chi(-\eta,\eta) - \chi(A_{i,-},A_{i,+}) \right)   \right .   \nonumber\\
   & &  \left .  + \frac{1}{2}A_{i,+} e^{-A^2_{i,-}} - \frac{1}{2}A_{i,-} e^{-A^2_{i,+}} - \eta e^{-\eta^2}  \right].
\label{eq:capture}
\end{eqnarray}

Here $\rho_{\rm DM} =0.3$~GeV/cm$^3$ is the local DM density, $\sigma_i$ is
the scattering cross section of DM on nuclei of type $i$ (its relation to
$\sigma_n$ will be specified in the next section), the dimensionless velocity $\eta = \frac{3}{2} \frac{u^2}{v_d^2}$, $v_e(r)$ is the
escape velocity from radius $r$, with
\[
\chi(a,b) = \frac{\sqrt{\pi}}{2} \biggl({\rm Erf(b)} - {\rm Erf(a)} \biggr),
\]
\[
A^2_i = 6 {v_e^2\over v_d^2} {m\, m_i\over (m-m_i)^2} \qquad {\rm and} \qquad A_{i,\pm} = A_i \pm \eta.
\]
The last factor in Eq.~(\ref{eq:capture}) is responsible for a
composition-dependent resonance-like behavior of the capture rate. Note that neglecting the motion of the Moon overestimates the capture rate by approximately factor four.

Before we proceed to estimate the evaporation and annihilation rates we have
to determine the distribution of DM particles in the Moon once they are
thermalized. To this end we compute the DM distribution semi-analytically
using methods described in
~\cite{Gould:1987ju,Gould:1989tu,Garani:2017jcj}. For $ \sigma_n <
\unit[10^{-33}]{cm^2}$, i.e. the so called optically thin regime, DM
distribution is isothermal given by
  \begin{equation}
    n_{\chi}(r) = N B e^{-{m \phi(r)\over T_\chi}}
    \label{eq:iso}
  \end{equation}
  with the normalization constant 
\[
B^{-1} = 4 \pi \int^{r_{\rm tidal}}_0 r^2
\dd r e^{-\frac{m \phi(r)}{T_\chi}},
\]        
where $\phi(r)$ is the gravitational potential of the Moon,  
$N$ is the total population of DM, and
$r_{\rm tidal}\sim 6\times 10^{4}{\rm km} \sim 35 R_{\leftmoon}$
is the tidal radius of Moon in the Earth gravitational field
\cite{Read:2005zm}. Assuming
Maxwell-Boltzmann velocity distribution for both DM and lunar matter, 
the DM temperature $T_\chi$ is determined by imposing that there is no net heat
transferred between the DM and lunar
matter~\cite{Spergel:1984re,Garani:2017jcj}. For DM masses $m
\gtrsim\unit[50]{GeV}$, we find that $T_\chi$ is equal to the core Moon
temperature ($T_{\rm core}$). However, for smaller masses $T_\chi$
asymptotically reaches $0.7 \times T_{\rm core}$.
For $\sigma >\unit[10^{-33}]{cm^2}$ the DM interacts with lunar
matter more than once per crossing, hence we assume that they are in local
thermodynamic equilibrium. In this case the DM distribution is again given by
Eq.~(\ref{eq:iso}) where $T_\chi$ should be replaced by $T_{\rm core}$. This
naive replacement is strictly true only if we consider the Moon to be an
isothermal sphere with temperature $T=T_{\rm core}= \unit[1700]{K}$,
i.e. there are no temperature gradients, otherwise the general expression
given in~\cite{Gould:1987ju,Gould:1989tu,Garani:2017jcj} should be utilized.
  
Having determined the DM profile in/around the Moon, we turn to the
annihilation and evaporation rates.  The coefficient $A$ in the annihilation
rate, the second term in Eq.~(\ref{eq:equilibrium}), for the s-wave
annihilation is given by
\begin{equation}
\label{eq:annihilationratesim}
A =  \langle \sigma_A v\rangle \, \frac{ \int n_\chi^2\,
  dV }{\left(\int n_\chi \, dV \right)^2}
\simeq  
\frac{ \langle \sigma_A v\rangle}{V_{th}}.
\end{equation}
Here $v$ is the relative velocity of the DM
particles and $V_{th} = 4/3 \pi R^3_{th}$ is the thermal volume, where the
thermal radius $R_{th} = (9 T_{\rm core}/(4 \pi G \rho_{\rm core} m))^{1/2} = 5.6\times 10^{6}
\,{\rm m} \left(\text{GeV}/m\right)^{1/2}$. The last (approximate) equality is
relevant in the case when the DM cloud is inside the Moon.
When presenting the results in the
next section, we use the exact expression Eq.~(\ref{eq:annrate-full})
rather than its approximation in Eq.~(\ref{eq:annihilationratesim}). 

We are now in a position to check whether the stationary regime is reached
during the Moon lifetime.  For $\sigma_n \gtrsim 10^{-33}\,{\rm cm^2}$, we can
make use of the geometric estimate for the capture rate ($C\sim 10^{24}\, {\rm
  s}^{-1} \left( {\rm GeV}/m\right)$) in order to find the condition on the
annihilation cross section that ensures that the stationary regime of DM
capture and annihilation is reached in less that $\tau \lesssim 1$~Gyr,
\[
\langle \sigma_A v\rangle \gtrsim 10^{-30} {\rm cm}^3/{\rm s}
\left({{\rm Gyr}\over \tau}\right)^2
\left({{\rm GeV}\over m}\right)^{1/2}.
\]
The standard thermal DM production cross section satisfies this condition. For
$\sigma_n \lesssim 10^{-39}\,{\rm cm^2}$, capture and annihilation are not in
equilibrium. Thus, neutrino telescope constraints on DM from annihilation in the Moon is not competitive with that from the Sun.

Finally, the evaporation rate, the last term in Eq.~(\ref{eq:equilibrium}) is
given by
\begin{eqnarray}
\label{eq:evap}
F &=& \sum_i \int_0^{R} s(r) \, \frac{n_\chi(r,t)}{N(t)} \,
4\pi r^2 \, \dd r \int_0^{v_e(r)} f_\chi(w, r)
\nonumber \\
  && \times 4 \pi w^2\, \dd w \int_{v_e(r)}^{\infty} R_i^+ (w \rightarrow v) \dd v ~.
\end{eqnarray}
where the factor $s(r)$ accounts for the suppression of the fraction of DM
particles that, even after acquiring a velocity larger than the escape
velocity, would actually escape from the Sun due to further interactions on
their way out, and is written as $s(r) = \eta_{\rm ang}(r) \, \eta_{\rm
  mult}(r) \, e^{-\tau(r)}$ ~\cite{Gould:1989tu}. Where $\tau(r) = \int_r^{R}
\ell^{-1}(r') \, \dd r'$ is the optical depth at radius $r$. The factors
$\eta_{\rm ang} (r)$ and $\eta_{\rm mult}(r)$ that take into account that DM
particles travel in non-radial trajectories and that multiple scatterings are
possible, are described in Appendix C of
Ref.~\cite{Garani:2017jcj}. The DM velocity distribution $f_\chi(w, r)$ is given by~Eq.~(\ref{eq:dist-full}), which is obtained by means of a full simulation 
described in appendix~\ref{App:dist}. The function $R_i^+ (w \rightarrow v)$ is the rate at which DM with velocity $w$ up-scatters to velocity $v > w$ upon collision with
nuclei~\cite{Gould:1987ju,Gould:1989tu,Garani:2017jcj}. In presenting the
results we evaluate Eq.~(\ref{eq:evap}) numerically.

To analytically understand what the evaporation mass could be, we can
conservatively neglect the annihilation term in Eq.~(\ref{eq:equilibrium}) and
examine the time evolution of the number of DM particles. Then, $N(t) = C/F (1
- e^{-F t})$.  We see that for evaporation to be negligible today we require
$F < 1/t_{\leftmoon}$, i.e. $F <\unit[3 \times 10^{-17}]{s^{-1}}$. In the
optically thin regime ($s(r)=1$), in the limit of equal DM and nuclei
temperature, and assuming constant nuclei density, the evaporation rate
becomes~\cite{Gould:1989tu}
\begin{eqnarray}
  F &=& \sum_i\int_0^{R}  \frac{n_\chi(r,t)}{N(t)} \,
4\pi r^2 \, \dd r \times  \nonumber  \\
&& \frac{2}{\sqrt{\pi}} n_i \sigma_i \left(\frac{2 T}{m}\right)^{1/2} e^{-\frac{m\,v_e^2}{2 T}}\left(\frac{m\,v_e^2}{2 T} - \frac{m_i}{2 m}\right).
  \end{eqnarray}
For example, when the target nuclei are Fe we find that
$m\gtrsim\unit[40]{GeV}$ for evaporation to be negligible for $\sigma_i = \unit[10^{-29}]{cm^2}$. Note that the maximal DM mass for evaporation is approximately set by the mass of the most abundant heaviest element DM scatters on. As illustrated in Sec.~\ref{sec:moon-heat-flow}, the core composition is dominated by $\rm{Fe}$ and the mantle is dominated by $\rm{O}$. Consequently the maximal DM mass for evaporation is expected to be in the range $m_{\rm O} < m < m_{\rm Fe}$.

{\it Moon Limb:} As the Moon does not have an atmosphere, we conservatively define the limb of the Moon to be $r_{l} = 3 \times R_{\leftmoon}$. The fraction of DM annihilating in the Moon limb is given by
\begin{equation}\label{eq:annfrac}
  \xi(m) =
  \frac{\int_R^{r_{l}} r^2 n^2_\chi(r,t)}{\int_R^{r_{\rm tidal}}
    r^2 n^2_\chi(r,t) + \int_0^R r^2 n^2_\chi(r,t)},
\end{equation}
here $n_\chi$ is obtained by integrating Eq.~(\ref{eq:dist-full}) over the velocity space. Note that $\xi(m<60\,{\rm GeV}) \neq 0$ due to highly eccentric orbits which are anisotropic. For example, we find $\xi(m=40\,{\rm GeV}) \approx 6 \times 10^{-3}$. Note also that tidal stripping of DM bound to the Moon due to the
gravitational field of the Earth is only important if DM thermal radius
extends beyond $\sim \unit[35]{R_{\leftmoon}}$, in which case the evaporation is too efficient for any constraints to be placed.

\section{Results}
\label{sec:results}

\begin{figure*}[!t]
\centering
\includegraphics[width=0.48\linewidth]{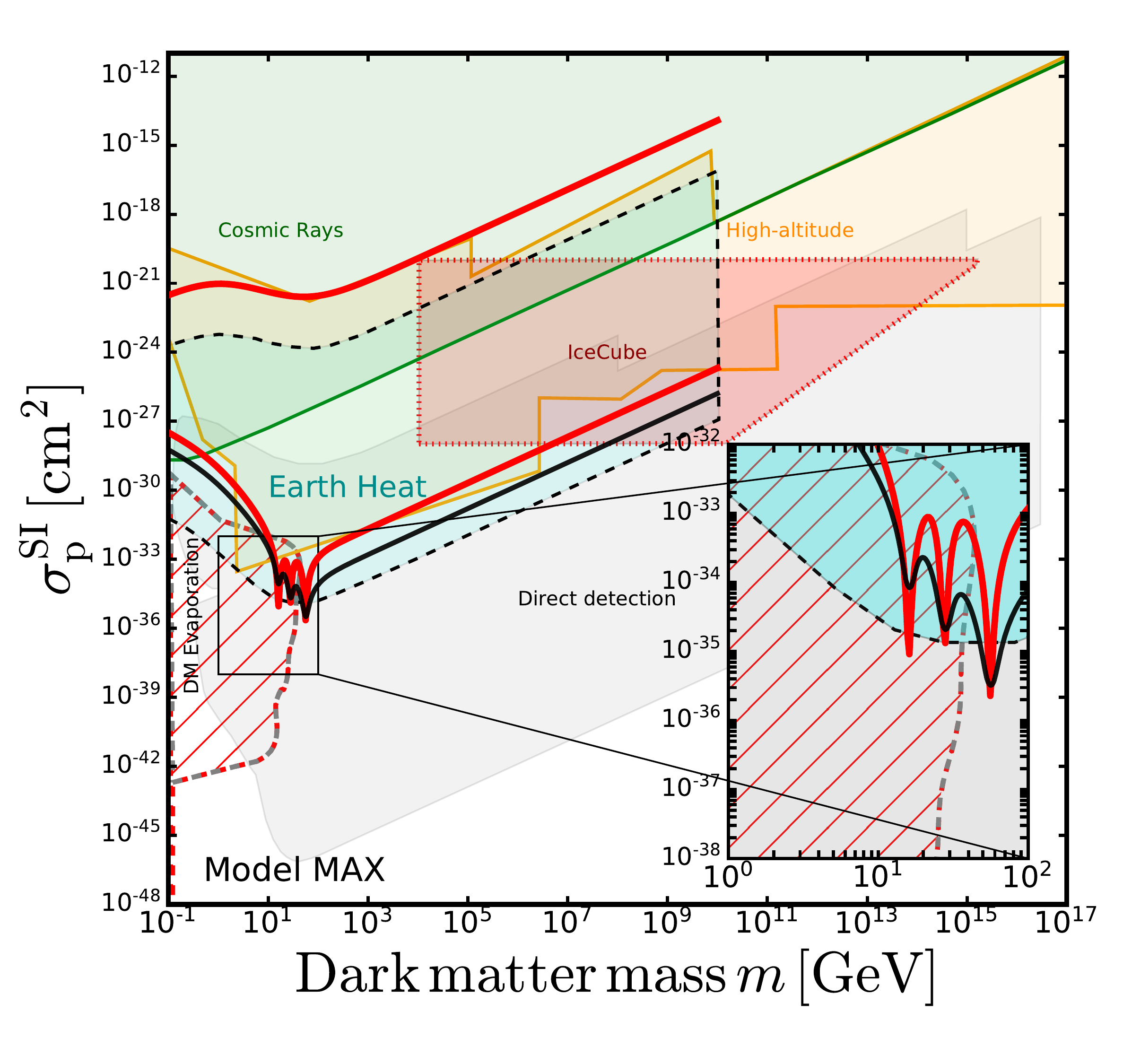}  \includegraphics[width=0.48\linewidth]{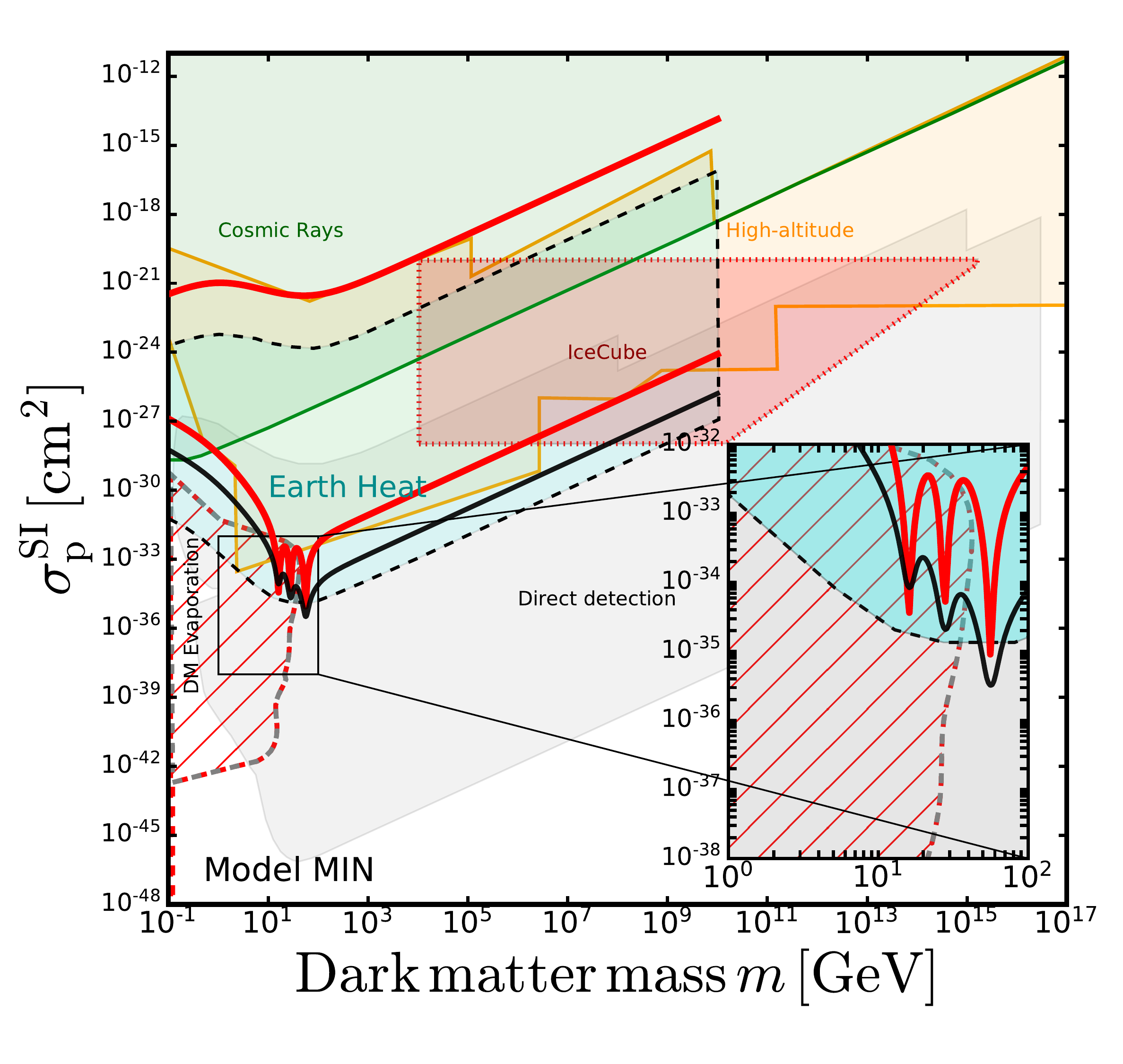} \\
\includegraphics[width=0.48\linewidth]{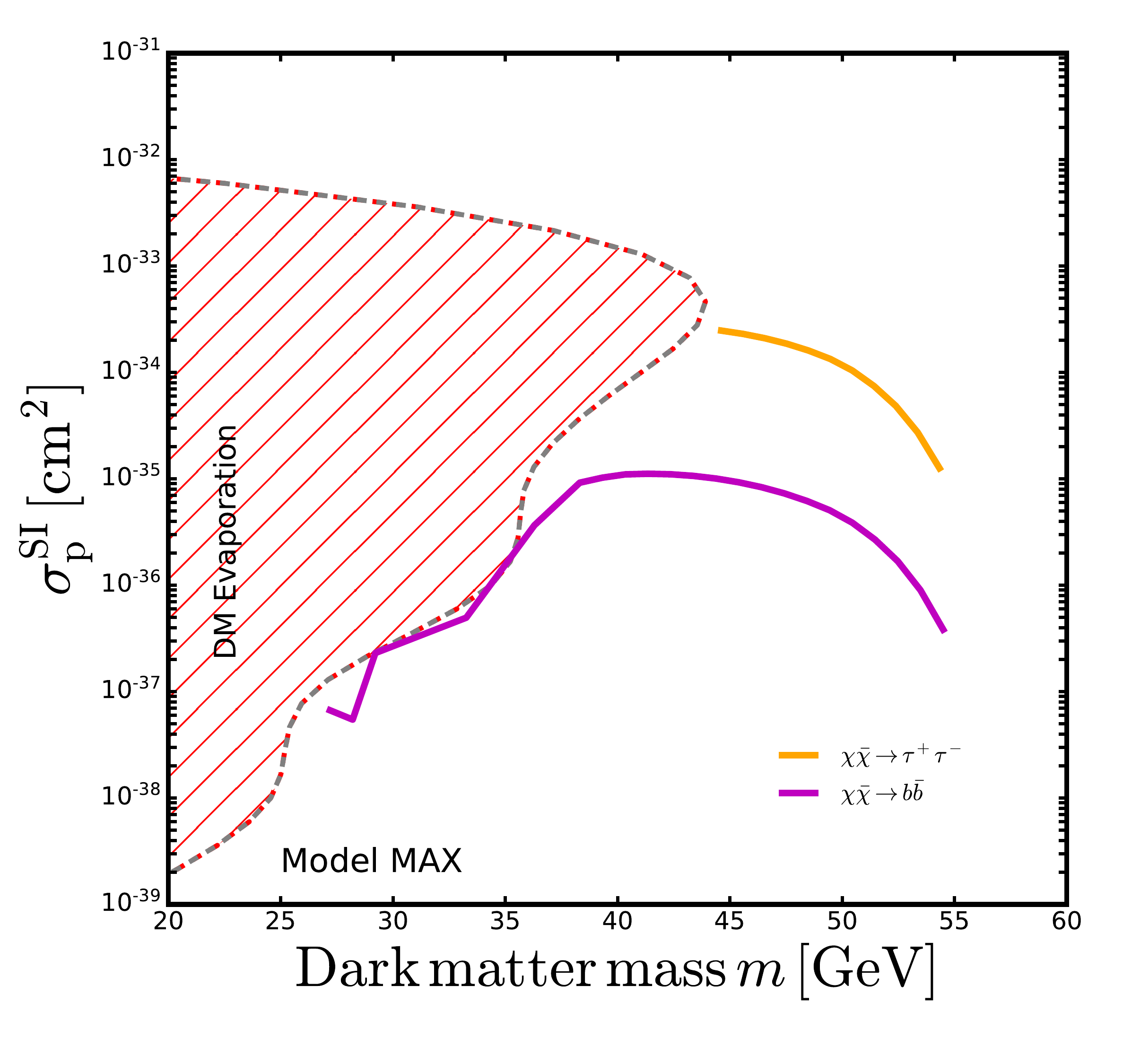}  \includegraphics[width=0.48\linewidth]{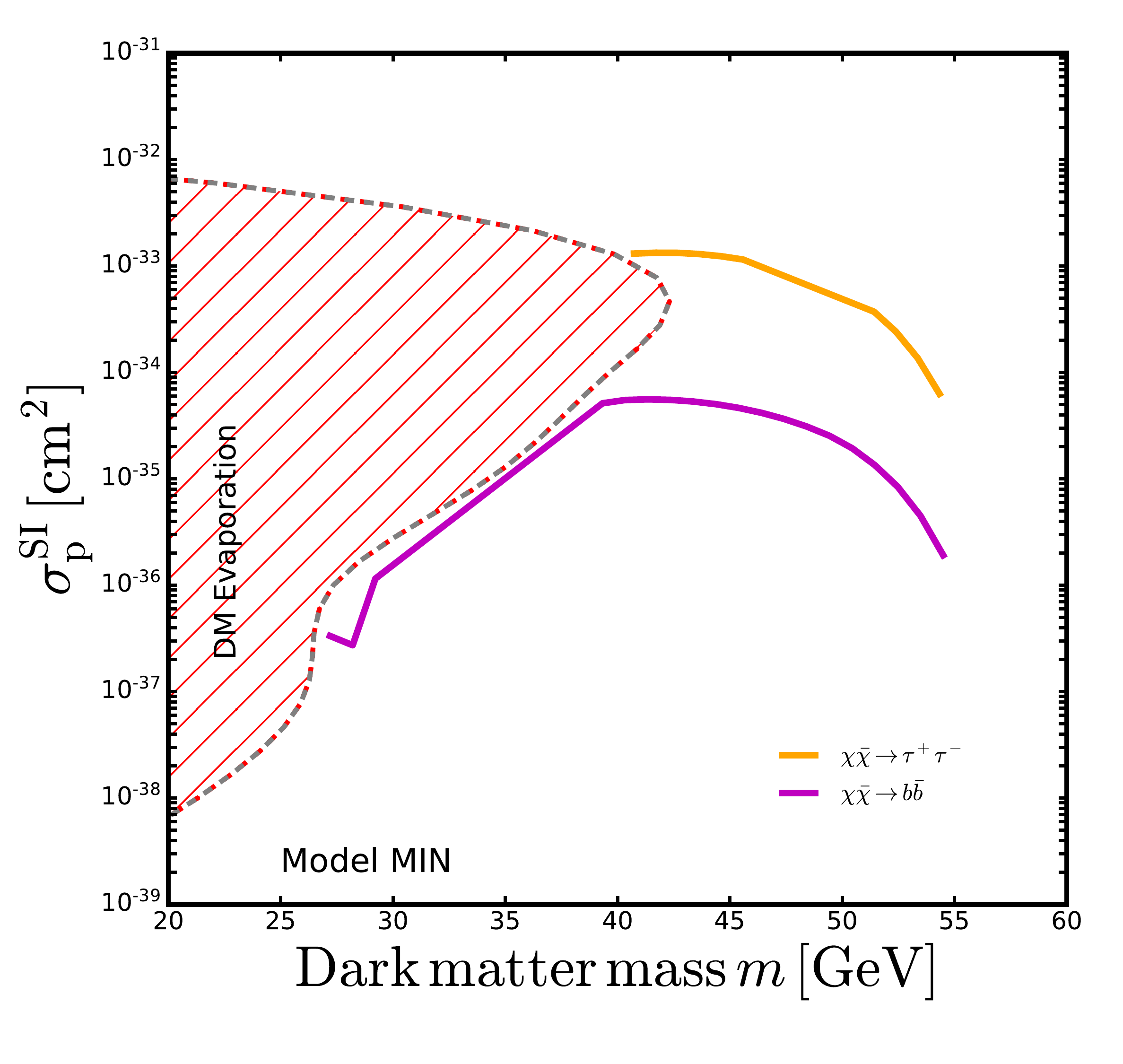}

	\caption{\textbf{\textit{Upper panels, exclusion curve from Moon heating for Moon reference models MAX (left) and MIN (right)}:}
The solid grey region shows constraints from surface and underground direct detection experiments~\cite{Davis:2017noy,Angloher:2017sxg,Petricca:2017zdp,Aprile:2017iyp,Kavanagh:2017cru}. The orange region (solid) is excluded by high-altitude experiments~\cite{Rich:1987st,Zaharijas:2004jv,Erickcek:2007jv,Barwick:1990hf,1978ApJ...220..719S}. Complementary constraints from IceCube are shown in red (dotted) regions~\cite{Albuquerque:2010bt}. Exclusion curve from the measurement of Earths' heat flux is shown in cyan (dashed)~\cite{Mack:2007xj}, and from DM-cosmic ray interactions in green (dotted)~\cite{PhysRevD.65.123503}. The above results were originally compiled in~\cite{Mack:2007xj,Kavanagh:2017cru}. Our results from heating of the Moon are shown in solid red and from heating of the Earth is shown in black. The red hatched regions show where DM evaporation from the Moon dominates over annihilation, thus no constraints hold in this region. \textbf{\textit{Lower panels, exclusion curve from DM annihilations in the Moon limb for Moon reference models MAX (left) and MIN (right):}} The exclusion curve for DM annihilation to $\bar{b}b$ is shown in purple. In orange the exclusion curve for DM annihilation to $\tau^+\tau^-$ is shown, see text for details. In the red hatched regions DM evaporation from the Moon dominates over annihilation.}
	\label{fig:exclusion}
\end{figure*}

To be specific, in the following we consider the spin-independent elastic
scattering cross section of DM on nucleons, $\sigma_n = \sigma_p^{\rm SI}$.
For a nucleus N with atomic mass $A_i$ and charge $Z_i$, assuming isospin
equivalence, the DM-nucleus spin-independent elastic scattering cross section
is
\begin{equation}\label{eq:si-xs}
\sigma_i^{\rm SI} = \left(\frac{\mu^{A_i}_r}{\mu^{p}_r}  \right)^2 A_i^2 \sigma^{\rm SI}_p.
\end{equation}
$\mu^{A_i}_r,\,\mu^{p}_r$ are reduced mass of nucleus-DM and proton-DM
respectively. Inserting this cross section in Eq.~(\ref{eq:capture}) we
evaluate the capture rate. As mentioned before DM heating of the Moon is given by $m\, C$. In the upper panels of Fig.~\ref{fig:exclusion} we present the exclusion curves (solid red) resulting from requirement that heating due to DM
annihilations in the Moon should not exceed the measured value of heat flow of
12 mW/m$^2$, for reference Moon models MAX (left) and MIN (right). We assume the DM annihilation rate to be $3\times 10^{-26}$ cm$^3$/s, which corresponds to the value for a thermal relic. We also assume that all the decay products of DM annihilation decay inside the Moon. For comparison we also present existing constraints on SI DM-proton cross section obtained from various other probes, in particular from DM heating of Earth (solid black). In the red hatched regions the captured DM
is efficiently evaporated from the Moon and constraints from heating do not
hold. Interestingly, we find that
the maximum evaporation DM mass is $\sim$ 40 GeV for $\sigma^{\rm SI}_p \sim 10^{-33}$ cm$^2$, for both the models MAX and MIN. Similarly we find that the DM evaporation mass for the Earth is $\sim$ 10 GeV for $\sigma^{\rm SI}_p \sim 10^{-34}$ cm$^2$. The critical cross section (i.e. the geometric cross section per nuclei) for the Moon (Earth) is about $10^{-32}$ cm$^2$ ($10^{-33}$ cm$^2$). For cross sections below that value, DM accretion is not efficient and crucially depends on the kinematics of scattering, resulting in 'resonance-like' enhancements as noted in~\cite{Gould:1987ir}. For the Moon, the most prominent peaks appear when DM mass matches that of the most abundant targets, which are iron, oxygen and silicon, respectively. Finally, these constraints do not extend infinitely to very large cross sections. Similar to the case of the Earth, for $\sigma^{\rm SI}_p \gtrsim 10^{-23}$ cm$^2$ DM gets stuck close to the surface thereby occupying a larger volume which suppresses the annihilation rate. However, we can still exclude some regions of parameter space with large cross sections that cannot be accessed by direct detection experiments. We evaluate this maximum cross section that could be constrained by balancing graviational force and drag force~\cite{Starkman:1990nj,Gould:1989gw}. We obtain the upper limit on the cross section by imposing DM particles at the surface drifts and settles within the thermal radius ($R_{th}$) in 4 Gyrs.
Furthermore, similar constraints can be placed on the spin dependent DM proton cross section. The resulting bounds can be obtained by an appropriate rescaling of Eq.~(\ref{eq:si-xs}). 

As mentioned before the DM thermal radius can be comparable to that of the Moon for DM mass $\sim \mathcal{O}(10)$ GeV, in which case some of the DM may annihilate just outside the moon forming a limb. The differential flux of $\gamma$-rays from annihilations in the limb is then given by
\begin{equation}\label{eq:limb-flux}
E^2\frac{\dd \Phi_\gamma}{\dd E} = \frac{ \Gamma}{4 \pi d^2} \xi(m) \sum_i {\rm{Br_i}} E^2 \frac{\dd N_\gamma}{\dd E},
\end{equation}
 
with $\xi(m)$ defined in Eq.~(\ref{eq:annfrac}). Here, $d= 3.8 \times 10^5$ km is the distance between the Moon and the Earth, $\Gamma= 1/2 A N^2$ is the total annihilation rate and $\dd N_\gamma/\dd E$ is the photon spectrum arising from a given final state, which we compute using Pythia~\cite{Sjostrand:2014zea}. 

The Fermi-large area telescope has measured the gamma-ray spectrum from the direction of the Moon~\cite{Cerutti:2016gts}. A maximum value of the differential photon flux of $ \approx 10^{-4}$MeV/cm$^2$s at $E \approx 100$ MeV is reported. The measured flux is consistent with gamma-ray production from cosmic ray collisions with Moon surface~\cite{Cerutti:2016gts}. For a given DM mass we obtain constraints on the cross section by requiring that the photon flux from DM annihilations,  Eq.~(\ref{eq:limb-flux}), does not exceed  $10^{-4}$ MeV/cm$^2$s at $E \approx 100$ MeV. We have also checked that the results do not significantly change upon comparison of the respective integrated photon fluxes. 

In the lower panels of Fig.~\ref{fig:exclusion} we present the exclusion curves resulting from DM annihilations in the Moon limb, for reference Moon models MAX (left) and MIN (right). The purple curve correspond to the case when DM annihilates only to $\bar{b} b$ and the orange curve to $\tau^+ \tau^-$, respectively.

Finally, comments are in order regarding possible constraints
that could arise from DM annihilations in the Earth
and Sun limbs. The approximate mass range where limb
signals are important is few times the maximal evaporation
mass for the assumed astrophysical object. As already mentioned,
the evaporation mass for the Earth is $\sim$ 10 GeV, so the
annihilations in the Earth limb are maximized for DM in the
mass range $10-20$ GeV. At heavier masses the annihilation fraction $\xi$,
 Eq.~(\ref{eq:annfrac}), 
vanishes exponentially. In the case of the Sun, it is well known that the
evaporation mass is about 4 GeV for SI interactions. Therefore,
DM annihilations in the solar limb could be important
in the mass range $4-10$ GeV. However, backgrounds from
cosmic ray interactions with the solar chromosphere could
pose a significant challenge~\cite{Seckel:1991ffa}. A precise determination
of the above constraints are left for future work.

It is important to note that several studies have examined
gamma-rays from DM annihilations in the halo around the
Sun~\cite{Strausz:1998my,Hooper:2001ij,Fleysher:2003iya,Atkins:2004qr,Sivertsson:2009nx}. Our analysis of the Moon limb differs in one main aspect:
we consider the DM population which has thermalized with the celestial
object. This population {\em accumulates until the capture rate balances the annihilation rate}.
Requiring that a sizable fraction of these annihilations
happens in the limb outside the object restricts
the DM mass range to few times the evaporation mass.
However, earlier works~\cite{Strausz:1998my,Hooper:2001ij,Fleysher:2003iya,Atkins:2004qr,Sivertsson:2009nx} considered relatively heavy
DM, i.e. 100 GeV or greater. Even though they consider
bound orbits, DM particles are not necessarily thermalized and are not efficiently accumulated on these orbits;
leading to a flux which could be extremely small. Consequently, the main result of the monte-carlo simulations of
annihilations in the halo~\cite{Sivertsson:2009nx} is that such limb observations cannot constrain DM properties for DM masses above 100 GeV.

\section{Conclusion}
\label{sec:concl}

We have demonstrated that complimentary constraints can be placed on DM elastic scattering cross section on protons using the available lunar data. Using the fact that the measured internal heat flux of Moon is 12 mW/m$^2$ we have derived constraints on thermally produced DM by requiring that annihilations from the captured DM should not overshoot this value. Constraints obtained in this way are competitive with similar constraints obtained from the Earth when DM mass matches with that of the target nuclei. We have also shown that for DM in the mass range $\sim$30 to 50~GeV the thermal radius of the DM cloud accumulated in the Moon is comparable to $R_{\leftmoon}$ and equilibrium between capture and annihilation can be achieved. In this case a small fraction $\sim 10^{-3}$ of DM annihilates just outside the Moon, leading to potentially observable $\gamma$-flux. We have obtained indirect constraints on DM parameters by comparing this flux with that of the Fermi-LAT measurement of $10^{-4}$ MeV/cm$^2$s at $E \approx 100$ MeV. 

The constraints derived in this paper are likely to be improved in the future. 
Several missions to the Moon are expected in the coming decade. New data from such exploration should reduce the current uncertainties on lunar composition, internal heat flow and temperature profile. This is likely to improve the constraints based on the Moon heat flux. Moreover, better understanding of the Moon temperature profile would allow us to more accurately determine the distribution of DM particles gravitaionally bound to the Moon. Consequently, the determination of the DM evaporation mass (i.e. the minimum DM mass for which evaporation would be negligible) and the constraints based on DM annihilations in the Moon limb would be improved.

Another potential improvement may be expected from a more detailed analysis of the $\gamma$-ray flux from the Moon limb that fits the space and energy distribution of produced photons to the Fermi-LAT data. We leave this for future work.

\section*{Acknowledgments}

We thank Sergio Palomares-Ruiz for discussions and comments.
RG is supported by the ULB-ARC grant ``Probing DM with Neutrinos'' and the Excellence of Science grant (EOS) convention~30820817. PT is supported by the IISN grant 4.4501.18.

\bibliographystyle{unsrt}
\bibliography{biblio}

\appendix

\section{DM distribution and the Moon Limb}
\label{App:dist}
The DM velocity distribution is an important input which is required to evaluate the DM evaporation and annihilation rates accurately. We use Monte Carlo methods discussed in Refs.~\cite{Gould:1989tu,Liang:2016yjf,Blennow:2018xwu} to compute the velocity distribution of DM particles gravitationally bound to the Moon.

The total energy of a particle moving in a central potential is 
\begin{equation}
m E = \frac{1}{2}m \dot{r}^2 + \frac{m}{2 r^2} J^2 + m \phi(r),
\end{equation}
where the gravitational potential is given by\\
$\phi(r) = G \int M(r')/r'^2 \dd r'\, \theta(R-r')   + GM/r \,\theta(r-R)$.

 For a given angular momentum $J$, the minimum energy of a particle whose trajectory intersects the Moon is 

\begin{equation}
E_{\rm min}(J) = {\rm Min}_{r \leq R} \frac{1}{2 r^2} J^2 + \phi(r) \equiv {\rm Min}_{r \leq R} V(J,r)
\end{equation}

We setup a simulation by distributing particles on a grid ($10^3 \times 10^3$) discretised in orbital parameters E and J. Denoting $f_{\alpha} \equiv f_{E,J}$ the number of particles in a state $\alpha$, its evolution is governed by the equation

\begin{equation}\label{eq:distfunc}
\dot{f}_{\alpha}=  C_\alpha  + \sum_{\beta} \mathcal{R}_{\alpha,\beta} f_\beta - f_\alpha \sum_\beta \Gamma_{\alpha,\beta} f_\beta .
\end{equation}
The first term above is the rate at which particles are captured in state $\alpha$, the last term is the annihilation term, while the second term is the rate of up-scatterings/down-scattering which includes the evaporation term~\cite{Blennow:2018xwu}
\begin{equation}
 \mathcal{R}_{\alpha,\beta}(r) = R_\beta(r) \frac{T(r_i, r_o)}{T(r_-, r_+)} \mathcal{P}_{\beta \rightarrow \alpha}(r).
\end{equation}
For trajectories that extend beyond the radius of the Moon we analytically match them to keplerian orbits.

The time spent by a particle in a shell between radius $r_1$ and $r_2$ is 

\begin{equation}
T(r_1,r_2) = \int^{r_2}_{r_1} \frac{\dd r}{ \dot{r}} =  \int^{r_2}_{r_1} \dd r \left(2(E - V(J,r)) \right)^{-1/2}
\end{equation}

For convenience Eq.~(\ref{eq:distfunc}) is solved ignoring the last term.
The radial distribution of DM is calculated by distributing particles of each state into all possible radii weighted by the fractional time spent at that radii 

\begin{equation}\label{eq:dist-full}
f(r) = \sum_\alpha f_\alpha \frac{T(r_i, r_o)}{T(r_-, r_+)}
\end{equation}

The total annihilation rate is given by
\begin{equation}
\label{eq:annrate-full}
A = \int \sigma_A|v_{\rm rel}| f(r,v_1)f(r,v_2) \dd^3 v_1 \dd^3 v_2 \dd^3 r,
\end{equation}
which is evaluated numerically.

\end{document}